\documentclass[12pt]{article}
\usepackage{graphicx}
\usepackage{subfigure}
\usepackage{amsmath}
\usepackage{amssymb}

\begin{document}

\begin{center}
\begin{large}
\title\\{ \textbf{A Potential Model Approach in the Study of Static and Dynamic properties of Heavy-Light Quark-Antiquark Systems.}}\\\

\end{large}

\author\

\textbf{$Sabyasachi\;Roy^{\emph{1}}\;and\:D.\:K.\:Choudhury^{\emph{2,3}}$ } \\\
\footnotetext{Corresponding author. e-mail :  \emph{sroy.phys@gmail.com}}
\textbf{1}. Department of Physics, Karimganj College, Karimganj, India.\\
\textbf{2}. Department of Physics, Gauhati University, Guwahati-781014, India.\\
\textbf{3}. Physics Academy of The North East, Guwahati-781014, India. \\

\begin{abstract}
We report some approximate analytic form of meson wave function constructed upon solving Schrodinger equation with linear plus Coulomb type Cornell potential. With this wave function, we study Isgur-Wise function and its derivatives for heavy-light mesons in the infinite heavy quark mass limit. We also explore the elastic form factors, charge radii and decay constants of pseudoscalar mesons in this QCD inspired quark model approach.  \\
\end{abstract}
\end{center}

\section{Introduction:}\rm
In the infrared energy region of QCD theory, potential model approach [1], although less fundamental than lattice QCD[2] or QCD sum rule[3], has proved to be successful even in the non-relativistic approximations, for the study of quark-antiquark bound states[4]. The use of non-relativistic model for heavy mesons is justified on the ground of large quark masses involved where velocities of heavy particles are non-relativistic. But, for mesons containing lighter quarks the validity of non-relativistic potential model approach depends mostly upon the choice of interaction potential. \\
There are actually several generally accepted potentials for modeling mesons. Potentials are generally constructed from the concepts of 'quark confinement' and 'asymptotic freedom'. In this regard, power law potentials are found to be very successful candidates. Some of the very generally accepted and commonly used potentials for the study of quark-antiquark bound states are mentioned below. \\
\begin{itemize}
  \item Cornell potential [5]: $V(r)=A+B r^{\alpha}$
  \item Martin potential [6] : $V(r)=A r - \frac{B}{r} +C $
  \item Logarithmic potential [7]: $V(r)=A+B \ln {r}$
  \item Song and Ling potential [8]:$ V(r) = A r^{1/2} +B r^{-1/2}$
  \item Turin potential [9]: $ V(r) = - A r^{-3/4} + B r^{3/4} +C $
  \item Richardson potential [10]: $V(r) = A r - \frac{B}{r \ln {\frac{1}{\bigwedge r}}}$
\end{itemize}

The basic condition in constructing these potentials are their flavour independence and existence of linear confinement. Of these, the Cornell potential is very well known phenomenological QCD motivated potential model. It is based on the two kinds of asymptotic behaviours - ultraviolet at short distance (Coulomb like) and infrared at large distance (linear confinement term). \\
In our present approach, we work with Cornell potential and develop wave function for heavy-light mesons by using non-relativistic Schrodinger equation [11] for bound state of its constituents. It is worthwhile to mention here that, getting exact solution of Schrodinger equation with such linear plus Coulombic potential has been the focus of interest for long. In atomic physics it corresponds to spherical Stark effect in hydrogen [12]. Several analytic and numerical techniques have been employed to get a reasonable solution of Schrodinger equation with such linear plus Coulombic potential. Here, first, we refer to some early work of H. Tezuka ( 1991) [13], where solution has been generated as some exponential function of interquark distance $r$. The solution, although, relatively simpler, has its own limitation. While extracting the analytic form of the solution, some additional 'counter terms' are incorporated in the potential function, which, in turn, has sacrificed the purity of the linear plus Coulombic nature of the potential. Recently, there has been some more attempts of solving Schrodinger equation with this linear plus Coulombic potential based on some rigorous quantum mechanical technique [14]. But, the analytic solution obtained there is not as such suitable for our study of hadron properties. \\
In this paper, we propose a simpler form of analytic solution of Schrodinger equation with Cornell potential. Our earlier work with Cornell potential using perturbation technique[15-17,72,73], considering its linear confinement term as parent, we have found that Airy's infinite polynomial function appears in the solution for wave function. Based upon our past study, here we construct the analytic form of the meson wave function for linear plus Coulombic potential in terms of Airy's function. To test the wave function we then study Isgur-Wise function (IWF)[18] of heavy-light mesons and its derivatives in the infinite heavy quark mass limit. The results for derivatives (slope and curvature) of IWF obtained by employing our proposed wave function matches reasonably well with recent theoretical and experimental results and other model predictions[19-26]. Then , we have explored different static and dynamic properties of mesons like electromagnetic form factor, charge radii, decay constants and made a comparative study with different theoretical and experimental expectations as far as practicable.\\
Here, it is to be made clear that, although non-relativistic potential models have been successful for heavy meson sector - still for mesons with one lighter constituent , its relativistic nature cannot be ignored in the study of properties of heavy-light mesons. Similar to our previous works [15-17,28-31, 72,73], here also we incorporate relativistic effect at the wave function level by introducing standard Dirac modification [32] in stead of full covariantization as in Bethe-Salpeter approach [33]. \\
With this introduction as section 1, we report the detailed formalism in section 2, results and calculations in section 3. We conclude by making our final comments in section 4. \\

\section{Formalism:}
\subsection{Wave function:}
The Cornell potential is of the standard form :
\begin{equation}
V (r) = -C_F \frac{\alpha_s}{r} + br + c
\end{equation}

$ C_F $ is the colour factor, which is given by :
\begin{equation}
C_F = \frac{N_C ^{2}-1}{2N_c}
\end{equation}
$ N_C $ is the colour quantum number; for $ N_C = 3 $, we have $ C_F = \frac{4}{3} $ and  Cornell potential takes the form :
\begin{equation}
V (r) = - \frac{4 \alpha_s}{3 r} + br + c
\end{equation}
For our convenience , we take $\frac{4 \alpha_s}{3 }=a$ so that our potential now becomes :
\begin{equation}
V (r) = - \frac{a}{r} + br + c
\end{equation}
Our Hamiltonian is ( considering $c=0$):
\begin{equation}
H = -\frac{\nabla^2}{2\mu}- \frac{a}{r}+br
\end{equation}
Here $\mu$ is the reduced mass of the meson with $m_1$ and $m_2$ as the individual quark masses.
\begin{equation}
\mu= \frac{m_1m_2}{m_1+m_2}
\end{equation}
The Schrodinger equation for the Hamiltonian $ H $ is $H|\Psi>=E|\Psi>$, from which we develop the two-body radial Schrodinger equation  [34] in terms of radial wave function $R(r)$ , as:
\begin{equation}
[-\frac{1}{2\mu}(\frac{d^2}{dr^2}+\frac{2}{r}\frac{d}{dr}-\frac{l(l+1)}{r^2})- \frac{a}{r}+br]R(r)=ER(r)
\end{equation}
Confining our consideration for ground state wave function ($l=0$):
\begin{equation}
[\frac{d^2}{dr^2}+\frac{2}{r}\frac{d}{dr}+2\mu (E+ \frac{a}{r}-br)]R(r)=0
\end{equation}
Here we introduce, $U(r)=r R(r)$, so that equation (8) transforms to :
\begin{equation}
\frac{d^2 U(r)}{dr^2}= 2\mu (br -\frac{a}{r}-E)U(r)
\end{equation}
Now to extract the solution, we consider two extreme conditions. \\

\textbf{Case-I:} \\
We take $ r \rightarrow \infty $ , so that $1/r$ term vanishes and equation (9) becomes :
\begin{equation}
\frac{d^2 U(r)}{dr^2}= 2\mu (br-E)U(r)
\end{equation}
Solution of this equation comes out in terms of Airy's function [35] as :
\begin{equation}
U(r) \sim Ai[\varrho] = Ai[\varrho_1 r +\varrho_0]
\end{equation}
Here, $\varrho= \varrho_1 r + \varrho_0$, with    $ \varrho_1 = (2 \mu b )^{1/3}$ and $\varrho_0 = - (\frac{2\mu}{b^2})^{1/3} E $. $\varrho_0$ is the zero of the Airy's function ($Ai[\varrho_0]=0$) and are given by [36]:
\begin{equation}
\varrho_0 = - [\frac{3\pi(4n-1)}{8}]^{2/3}
\end{equation}

\textbf{Case-II:} \\
Now, if we take $ r \rightarrow 0 $ , then $1/r$ term in equation (9) will prevail:
\begin{equation}
\frac{d^2 U(r)}{dr^2}= 2\mu (-\frac{a}{r}-E)U(r)
\end{equation}
The solution of this equation is :
\begin{equation}
U(r) \sim e^{-r/a_0}
\end{equation}
Here, $a_0 = \frac{1}{\mu a}=\frac{3 }{4\mu \alpha_s}$.
We construct the purely analytic solution for ground state as the multiplication of the solutions of these two extreme cases:
\begin{equation}
U(r) \sim Ai[\varrho_1 r +\varrho_0] e^{-r/a_0}
\end{equation}
With $N$ as the normalisation factor, our radial wave function has thus the form :
\begin{equation}
\Psi(r) = \frac{N}{r} Ai[\varrho_1 r +\varrho_0] e^{-r/a_0}
\end{equation}
Considering relativistic effect on the wave function following Dirac modification, the relativistic wave function is given by:
\begin{equation}
\Psi_{rel} (r) = \frac{N}{r} Ai[\varrho_1 r +\varrho_0] e^{-r/a_0}(\frac{r}{a_0})^{-\epsilon }
\end{equation}
Now, $N$ is the normalisation constant for the relativistic wave function.
Here,
\begin{equation}
\epsilon = 1-\sqrt{1-(\frac{4\alpha_s}{3})^2}=1-\sqrt{1-a^2}
\end{equation}
Airy's infinite series as a function of $\varrho =\varrho_1 r +\varrho_0 $ can be expressed as [37,38] :
\begin{eqnarray}
Ai[\varrho_1 r + \varrho_0] = a_1[1+\frac{(\varrho_1 r + \varrho_0)^3}{6}+\frac{(\varrho_1 r + \varrho_0)^6}{180}+\frac{(\varrho_1 r + \varrho_0)^9}{12960}+...]- \nonumber \\
 b_1[(\varrho_1 r + \varrho_0) +\frac{(\varrho_1 r + \varrho_0)^4}{12}+\frac{(\varrho_1 r + \varrho_0)^7}{504}+\frac{(\varrho_1 r + \varrho_0)10}{45360}+...]
\end{eqnarray}
\begin{flushright}
 with $a_1=\frac{1}{3^{2/3} \Gamma(2/3)}=0.3550281$ and $ b_1=\frac{1}{3^{1/3} \Gamma(1/3)} =0.2588194.$ \\
\end{flushright}
We consider Airy's series up to $O(r^3)$:
\begin{equation}
Ai[\varrho] = a_1[1+\frac{(\varrho)^3}{6}]- b_1\varrho
\end{equation}
From this , we get the Airy function as an explicit function of $r$ as:
\begin{equation}
Ai[r] = k_0 + k_1 r + k_2 r^2 +k_3 r^3
\end{equation}
with $k_i s$ having their explicit form as given below:
\begin{eqnarray}
k_0 = a_1 + \frac{a_1 \varrho_0 ^3 }{6} - b_1 \varrho_0 \\
k_1 = \frac{a_1 \varrho_0^2 \varrho_1 }{2} -b_1 \varrho_1 \\
k_2 = \frac{a_1 \varrho_0 \varrho_1^2}{2} \\
k_3 = \frac{a_1 \varrho_1^3 }{6}
\end{eqnarray}
With this truncated expression of Airy's function, we now construct the wave function as :
\begin{eqnarray}
\Psi_{rel} (r) = \frac{N}{r} [k_0 + k_1 r + k_2 r^2 +k_3 r^3] e^{-r/a_0}(\frac{r}{a_0})^{-\epsilon }\\
= \frac{N}{a_0^{-\epsilon}}[ k_0 r^{-1-\epsilon} + k_1 r^{-\epsilon} + k_2 r^{1-\epsilon} +k_3 r^{2-\epsilon} ]e^{-r/a_0}
\end{eqnarray}

\subsection{Isgur-Wise Function:}
Under Heavy Quark Symmetry ( HQS ), the strong interactions of heavy quarks are independent of its spin and mass[39] and all the form factors are completely determined at all momentum transfers in terms of the universal IWF. It is useful to parameterize IWF in terms of its derivatives at zero recoil ( y=1)[40]. In explicit form, for small non-zero recoil, IWF can be expressed as:
\begin{equation}
\xi(y)=1-\rho^2 (y-1) +C(y-1)^2+\cdots\cdots
\end{equation}
Thus, HQS provides us with a prediction for the normalisation of the IWF at zero recoil point (y=1).
Here $\rho^2$ is the slope ( charge radii ) and $C$ is the curvature ( convexity parameter) of IWF, which are measured at zero recoil point as :
\begin{equation}
\rho^2 = -\frac{\delta\xi (y)}{\delta y}|_{y=1} \;\;,\;\; C=\frac{\delta^{2}\xi (y)}{\delta y^2}|_{y=1}
\end{equation}

It should be mentioned here that for the reliable analysis of the IWF, the first two terms in the expansion of IWF (equation (28)) are required to be taken into consideration, thus making it necessary to calculate both slope and curvature parameters. The calculation of $ \rho^{2}$ and $C $ provides a measure of the validity of HQET in infinite mass limit along with a valid test for confirmation of our wave function. There have been several attempts to calculate $ \rho^{2}$ and $C$ from theory and models[19-26]. The corresponding results are shown in Table-2. On general ground, the slope parameter should have value around unity and curvature of IWF is expected to have small positive value for all $y>1$. \\
The calculation of this IWF is non-perturbative in principle and is performed for different phenomenological wave functions of mesons [41]. This function depends upon the meson wave function and some kinematic factor, as given below :
\begin{equation}
\xi(y)=\int_0 ^\infty 4\pi r^2 |\Psi(r)|^2\cos(pr)dr
\end{equation}
where $\cos(pr)=1-\frac{p^2 r^2}{2}+\frac{p^4 r^4}{4}$ +$\cdot\cdot\cdot\cdot\cdot\cdot$  with $ p^2=2\mu^2 (y-1).$ Taking $cos(pr)$ up to  $O(r^4)$ we get,

\begin{equation}
\xi(y)= \int_0 ^\infty 4\pi r^2 |\Psi(r)|^2dr - [4\pi\mu^2\int_0^\infty r^4|\Psi(r)|^2dr](y-1)+[\frac{2}{3}\pi\mu^4\int_0^\infty r^6|\Psi(r)|^2dr](y-1)^2
\end{equation}
Equations (28) and (31) give us :

\begin{eqnarray}
\rho^2 = [4\pi\mu^2\int_0^\infty r^4|\Psi(r)|^2dr] \\
C= [\frac{2}{3}\pi\mu^4\int_0^\infty r^6|\Psi(r)|^2dr] \\
\int_0 ^\infty 4\pi r^2 |\Psi(r)|^2dr =1
\end{eqnarray}
FRom equation (34) we can calculate the normalization constants $ N \;\;$ for the wave function.\\

\subsection{Form factors and charge radii:}
In order to test the wave function further on, over and above IWF, we calculate the elastic charge form factors, decay widths and charge radii of pseudoscalar mesons. Here we define the elastic charge form factor for a charged system of point quarks [42,43]as:
\begin{equation}
eF(Q^2)=\sum_i\frac{e_i}{Q_i}\int_0^{\infty}r\mid \Psi(r) \mid^2 \sin(Q_i r )dr
\end{equation}
Applying equation (27) in the above equation (35),we obtain:
\begin{equation}
eF(Q^2)= N^2 a_0^{2\epsilon} \Sigma_i\frac{e_i}{Q_i}\int_0 ^{\infty} \sum_{l=1}^{7}C_l r^{P_l -1} e^{-a_2 r } \sin (Q_i r)dr
\end{equation}
with $a_2= \frac{1}{a_0}$ and $P_l = -(1+2 \epsilon -l )$. \\

Here, $C_l s$ are having the following expressions:
\begin{eqnarray}
C_1=k_0^2 \\
C_2=2 k_0 k_1 \\
C_3=k_1^2 + 2 k_0 k_2 \\
C_4=2 k_0 k_3 +2 k_1 k_2 \\
C_5=k_2 ^2 +2k_1 k_2 \\
C_6=2 k_2 K_3 \\
C_7= k_3^2
\end{eqnarray}
Equation (36) upon integration and further simplification [ shown in Appendix-A] , gives:
\begin{eqnarray}
eF(Q^2)= N^2 a_0^{2\epsilon} \Sigma_{l=1}^7 C_l \Gamma(P_l) P_l[e_1(a_2^2+Q_1^2)^{-\frac{P_l+1}{2}}+e_2(a_2^2+Q_2^2)^{-\frac{P_l+1}{2}}]\\
eF(Q^2)= N^2 a_0^{2\epsilon} \Sigma_{l=1}^7 C_l \Gamma(P_l) P_l (m_1 +M_2)^{P_l +1} (\frac{-P_l-1}{2})  \\\nonumber
[e_1{a_2^2(m_1+m_2)^2 +m_2^2 Q^2 }^(\frac{-P_l-3}{2}) m_2^2 + e_2{a_2^2(m_1+m_2)^2 +m_1^2 Q^2 }^(\frac{-P_l-3}{2}) m_1^2]
\end{eqnarray}
The fraction of virtuality carried by individual quark are:
\begin{eqnarray}
Q_1= \frac{m_2 Q}{m_1 + m_2}=\frac{\mu}{m_1}Q \\
Q_2= \frac{m_1 Q}{m_1 + m_2}=\frac{\mu}{m_2}Q
\end{eqnarray}

In the infinite heavy quark mass limit, $m_1\rightarrow\infty$ and $\mu \rightarrow m_2$, so that $Q_1 \rightarrow 0$ and $Q_2 \rightarrow Q$.
Under this infinite mass consideration, equation (44) transforms  into:
\begin{equation}
eF(Q^2)\mid_\infty= N^2 a_0^{2\epsilon} \Sigma_{l=1}^7 C_l \Gamma(P_l) P_l[e_1 a_2^{-\frac{P_l+1}{2}}+e_2(a_2^2+Q^2)^{-\frac{P_l+1}{2}}]
\end{equation}

The average charge radius square of mesons is obtained from:
\begin{equation}
<r^2>=-6\frac{d}{d Q^2}(eF(Q^2))|_{Q^2=0}\\
\end{equation}
Using equation (44) in this expression, we calculate the charge radii of mesons as[details shown in Appendix-A]:
\begin{equation}
<r^2> = 2^{2\epsilon-2}3 N^2 a_0^2 \frac{e_1 m_2^2 +e_2 m_1^2}{(m_1+m_2)^2}\sum_{l=1}^7 C_l \Gamma(P_l) P_l(P_l+1)a_0^l
\end{equation}
Under infinite heavy quark mass limit :
\begin{equation}
<r^2>_\infty = 2^{2\epsilon-2}3 N^2 a_0^2 e_2\sum_{l=1}^7 C_l \Gamma(P_l) P_l(P_l+1)a_0^l
\end{equation}
The relation between $<r^2>$ and $<r^2>_\infty$ is given by :
\begin{equation}
<r^2> =  <r^2>_\infty [\frac{e_1 m_2^2 +e_2 m_1^2}{e_2(m_1+m_2)^2}]
\end{equation}

\subsection{Decay Constants:}
$B$ and $D$ mesons can undergo a weak decay via annihilation of their constituent quarks. The rate of decay depends upon the CKM matrix elements and a number called the decay constant ( $f_B$ or $f_D$). The decay constant gives the probability with which the quark and antiquark are found inside the meson at the same point and can annihilate.
\begin{equation}
\Gamma \infty \mid V_{Qq}\mid f_{Qq}
\end{equation}
In our case of study, we take the standard expression (Van Royen-Weisskopf formula) of pseudoscalar meson decay constant [70] in non-relativistic quark model to be :
\begin{equation}
f_P = \sqrt{\frac{12}{M_P}|\Psi(0)|^2}
\end{equation}
Here, $M_P$ is the physical mass of the pseudoscalar meson, $\Psi(0)$ is the wave function of meson at the origin ( $r=0$). Thus, it can be said that a study of the decay constant of meson is equivalent to a study of its wave function at origin.
It is to be mentioned here that our relativistic wave function results divergence while calculating $\Psi_{rel}(0)$ due to the factor $r^{-\epsilon}$. As a result, we are bound to consider the non-relativistic version of equation (26), which on expansion gives,
\begin{equation}
\Psi(r) = N[k_0 r^{-1} + k_1 r + k_2 r^2 +k_3 r^3] [1+(\frac{-r}{a_0})+\frac{(\frac{-r}{a_0})^2}{2!}+\frac{(\frac{-r}{a_0})^3}{3!}+\cdots]
\end{equation}
From this we compute the wave function at origin as :
\begin{equation}
\Psi(0) = N[k_1-\frac{k_0}{a_0}]
\end{equation}
$k_0$ and $k_1$ are given by equations (22-23). Using equation (56) in (54), we can calculate decay constant of mesons.
\section{Calculation and result:}
With the constructed wave function (equation 27), we proceed to calculate meson properties. To start with, we study the IWF and its derivatives in the infinite heavy quark mass limit of heavy-light mesons. We confine within B and D sector mesons. The results of slope and curvature of heavy-light mesons are shown in Table-1. For comparison,  in Table-2, we report some standard experimental and theoretical results. The variation of IWF $\xi(y)$ vs $y$ is shown in Fig-1, which confirms that the boundary condition for zero recoil is satisfied throughout.

We have studied the variation of form factor values of $B$ and $D$ mesons with virtuality $Q^2$. The results are shown in Table-3 and 4, which are further plotted in Figure-2. While exploring variation of form factor at infinite heavy quark mass limit ( which are also shown in Table 3 and 4), we find that the our results do not vary much as regards form factor values with finite mass are concerned.\\

The values of average charge radii square for different $B$ and $D$ sector mesons are shown in Table 5. We have also calculated the ratio $R=\frac{<r^2>}{<r^2>_\infty}$ for each of these mesons which are also incorporated in Table 5. The plot of this ratio $R$ with mass of heavy quark is shown in Figure-3. \\

Lastly, following equation (56), we have also computed the decay widths of these mesons and the results are shown in Table 6. For comparison, the related results from different models and experiments are shown in Tables 7 , 8 and 9. \\

\section{Conclusion and remarks:}

We have developed an approximate analytic ground state wave function of meson with Coulomb plus linear potential. In analogy to our previous works [15-17], here also we have studied the IWF of heavy-light mesons with this wave function. Regarding slope and curvature of IWF ( Table-1), we find that our results becomes better for heavier mesons. Also, the graph of $\xi(y)$ vs $y$ (Fig-1) confirms that the zero recoil condition is maintained, confirming the validity of our model.\\

Then, we have further applied this wave function to calculate electromagnetic form factors, charge radii and decay widths of mesons. While estimating the validity of our result by comparison with theoretical and experimental results, we put forward the following observations. \\
\begin{itemize}
  \item Unlike as in [28,29], here in our calculation, we have worked with fixed values for confinement parameter $b$ ( 0.183 $GeV^2$) and coupling constant $\alpha_s$ ( 0.39 for $D$ meson and 0.22 for $B$ meson)[5]. This,in many cases, has reduced the flexibility of taking the results closer to expectations.
  \item Regarding our results for form factor, from Figure-2 we find that the form factors decreases with increase in $Q^2$ for both $D$ and $B$ mesons, as it should be. However, due to lack of experimental results for form factors for these sectors of mesons, we are handicapped in carrying out any comparative analysis. We make a note of the point that our results of form factors with infinite mass limit do not result in appreciable difference with that of finite mass limit. However, our results for form factor are definitely an improvement over that of [71].
  \item As charge radius follows as a consequence of form factor, we have also calculated the same for different $D$ and $B$ sector mesons. As the mean square charge radius for heavy pseudoscalar mesons have not been measured yet, we compare our results with some theoretical expectation [ Huang paper]. We find that our results are more or less in good agreement. The plot of $R=\frac{<r^2>}{<r^2>_\infty}$ with $Q^2$ (Figure- 3) suggests that the infinite mass limit is reasonable for charm quark but may be not for bottom quark.
  \item Regarding our calculation of decay constant from our constructed wave function, we are ought to make a naive confession that our calculation of form factor (equation) is relatively crude. This is because, the calculation of decay constant veers round the parameter `wave function at origin $\Psi(0)$' and to overcome divergence, we are compelled to make some approximation in its calculations. With such approximation, still our results are comparatively in good agreement with other expectations ( Table 7-9).  Also, instead of separately calculating the physical mass for mesons ( $M_P$ ) from some analytical expression ( which itself again involves $\Psi(0)$), here we have calculated decay constant with standard values of meson masses which are reported in Table -6.
\end{itemize}

Lastly, we conclude by making the following comment. \\
We identify our this venture as a new approach in finding wave function of meson when compared with our earlier works [15-17] following perturbation technique. Our approximate analytical wave function for mesons works reasonably well for the studies of static and dynamic properties of mesons within its limitations. Our wave function is for only ground state mesons, and as such there remains further scope for improvement of formalism for higher spectroscopic states.

\appendix

\numberwithin{equation}{section}

\begin{center}
\section{(Appendix)}
\end{center}
We have, from equation (27),
\begin{equation}
\Psi_{rel} (r)=\frac{N}{a_0^{-\epsilon}}[ k_0 r^{-1-\epsilon} + k_1 r^{-\epsilon} + k_2 r^{1-\epsilon} +k_3 r^{2-\epsilon} ]e^{-r/a_0}
\end{equation}
From this , we get :
\begin{eqnarray}
|\Psi_{rel} (r)|^2=\frac{N^2}{a_0^{-2\epsilon}}[ k_0^2 r^{-1-2\epsilon} +k_1^2 r^{1-2\epsilon}+k_2^2 r^{3-2\epsilon}+k_3^2 r^{5-2\epsilon}+2k_0k_1 r^{-2\epsilon}+ \\\nonumber
2k_0k_2 r^{1-2\epsilon}+2k_0k_3 r^{2-2\epsilon}+2k_1k_2 r^{2-2\epsilon}+2k_1k_3 r^{3-2\epsilon}+2k_2k_3 r^{4-2\epsilon}] e^{-2r/a_0}
\end{eqnarray}
We apply this in equation (35) to get :
\begin{equation}
eF(Q^2)=\frac{N^2}{a_0^{-2\epsilon}}\Sigma_i\frac{e_i}{Q_i}\sum_{l=1}^7 C_l I_l
\end{equation}
Here,
\begin{equation}
I_l=\int_0^\infty r^{P_l-1}e^{-a_2 r} sin(Q_i r)dr
\end{equation}
with $a_2=\frac{2}{a_0}$, $P_l=-(1-2\epsilon -l)$ and $C_l$ are given in equations (37-43).

This integration has the standard solution :
 \begin{equation}
I_l= \frac{\Gamma(P_l)sin(P_l\phi_i)}{(a_2^2+Q_i^2)^{P_l/2}}
\end{equation}
Here, we have taken ,
\begin{eqnarray}
\phi_i = \sin^{-1}\frac{Q_i}{(a_2^2+Q_i^2)^{1/2}}\\
\approx \frac{Q_i}{(a_2^2+Q_i^2)^{1/2}}\\
\sin(P_l \phi_i) \approx P_l \phi_i = \frac{P_l Q_i}{(a_2^2+Q_i^2)^{1/2}}
\end{eqnarray}

Now from equation (A.3), we have-

\begin{eqnarray}
eF(Q^2)= N^2 a_0^{2\epsilon} \Sigma_i\frac{e_i}{Q_i}\int_0 ^{\infty} \sum_{l=1}^{7}C_l r^{P_l -1} e^{-a_2 r } \sin (Q_i r)dr \\
=N^2 a_0^{2\epsilon} \Sigma_i\frac{e_i}{Q_i}\sum_{l=1}^{7}C_l\int_0 ^{\infty}r^{P_l -1} e^{-a_2 r } \sin (Q_i r)dr \\
=N^2 a_0^{2\epsilon} \Sigma_i\frac{e_i}{Q_i}\sum_{l=1}^{7}C_l\frac{\Gamma(P_l)\sin(P_l \phi_i)}{(a_2^2+Q_i^2)^{P_l/2}}\\
\end{eqnarray}
\begin{eqnarray}
=N^2 a_0^{2\epsilon} \Sigma_i\frac{e_i}{Q_i}\sum_{l=1}^{7}C_l\frac{\Gamma(P_l)P_l Q_i}{(a_2^2+Q_i^2)^{P_l/2}(a_2^2+Q_i^2)^{1/2}}\\
=N^2 a_0^{2\epsilon}\sum_{l=1}^{7}C_l\Gamma(P_l)P_l\Sigma_i\frac{e_i}{(a_2^2+Q_i^2)^{(P_l+1)/2}} \\
=N^2 a_0^{2\epsilon}\sum_{l=1}^{7}C_l\Gamma(P_l)P_l[e_1(a_2^2+Q_1^2)^{-(P_l+1)/2}+e_2(a_2^2+Q_2^2)^{-(P_l+1)/2}]\\
=N^2 a_0^{2\epsilon}\sum_{l=1}^{7}C_l\Gamma(P_l)P_l(m_1+m_2)^{P_l+1}[e_1\{a_2^2(m_1+m_2)^2+m_2^2Q^2\}^{-(P_l+1)/2} \\\nonumber
+e_2\{a_2^2(m_1+m_2)^2 +m_1^2Q^2\}^{-(P_l+1)/2}]
\end{eqnarray}
Here,
\begin{eqnarray}
\frac{P_l+1}{2}=\frac{-2\epsilon +l}{2}  \\
Q_1=\frac{m_2 Q}{m_1+m_2} \\
Q_2=\frac{m_1 Q}{m_1+m_2}
\end{eqnarray}
Now,
\begin{eqnarray}
\frac{d}{dQ^2}\{eF(Q^2)\}=N^2 a_0^{2\epsilon}\sum_{l=1}^{7}C_l\Gamma(P_l)P_l(m_1+m_2)^{P_l+1}(\frac{-P_l-1}{2}) \\\nonumber
[e_1\{a_2^2(m_1+m_2)^2+m_2^2Q^2\}^{(-P_l-3)/2}m_2^2 + e_2\{a_2^2(m_1+m_2)^2+m_1^2Q^2\}^{(-P_l-3)/2}m_1^2]
\end{eqnarray}
From this we compute the charge radius,
\begin{eqnarray}
<r^2>=-6\frac{d}{dQ^2}\{eF(Q^2)\}\\
=6 N^2 a_0^{2\epsilon}\sum_{l=1}^{7}C_l\Gamma(P_l)P_l(m_1+m_2)^{P_l+1}(\frac{P_l+1}{2}) \\\nonumber
[e_1\{a_2^2(m_1+m_2)^2+m_2^2Q^2\}^{(-P_l-3)/2}m_2^2 + e_2\{a_2^2(m_1+m_2)^2+m_1^2Q^2\}^{(-P_l-3)/2}m_1^2] \\
=3 N^2 a_0^{2\epsilon}\sum_{l=1}^{7}C_l\Gamma(P_l)P_l(m_1+m_2)^{P_l+1}(P_l+1)a_2^{-P_l-3}(m_1+m_2)^{-P_l-3}[e_1m_2^2+e_2m_1^2]\\
=2^{2\epsilon-2}3 N^2 a_0^2\frac{e_1m_2^2+e_2m_1^2}{(m_1+m_2)^2}\sum_{l=1}^{7}C_l\Gamma(P_l)P_l(P_l+1)a_0^l
\end{eqnarray}

\begin{center}
$*****$
\end{center}

\begin{table}[ht]
\begin{center}
\caption{Result of $\rho^2$ and $C$ with $\Psi_{rel}(r)$ .}\label{cross}
\begin{tabular}{|c|c|c|}
  \hline
  $mesons$  & $\rho^2$ & $C$  \\
  \hline \hline
  $D$    & 0.6681 & 0.1483    \\
  $B$    & 0.7688 & 0.1999    \\
  $D_s$  & 0.9095 & 0.2850    \\
  $B_s$  & 1.1039 & 0.4275    \\
  \hline
\end{tabular}
\end{center}
\end{table}

\begin{table}[ht]
\begin{center}
\caption{Results of slope and curvature of $\xi(y$) in different models and collaborations.}\label{cross}
\begin{tabular}{|c|r|r|}
  \hline
Model / collaboration &	Value of slope & Value of curvature \\
\hline \hline
 Ref [15] & 0.7936 & 0.0008 \\
 Le Youanc et al [19] & $\geq 0.75$ & $\geq 0.47$ \\
 Skryme Model [20] & 1.3 & 0.85 \\
 Neubert [21] & 0.82  0.09 & -- \\
 UK QCD Collab. [22]  & 0.83 & -- \\
 CLEO [23] & 1.67 & -- \\
 BELLE  [24] & 1.35 & -- \\
HFAG [25] & 1.17 $\pm 0.05$ & -- \\
Huang [26] & 1.35 $\pm 0.12 $ & -- \\
\hline
\end{tabular}
\end{center}
\end{table}

\begin{table}[ht]
\begin{center}
\caption{Form Factor Values of $D$ meson in finite and infinite mass limit.}\label{cross}
\begin{tabular}{|c|c|c||c|c|c|}
  \hline
 $Q^2$ & $F(Q^2)$ & $F(Q^2)|_{\infty}$ &  $Q^2$ & $F(Q^2)$& $F(Q^2)|_{\infty}$  \\
\hline \hline
0    &  1     & 1       & 0.8 & 0.1368 & 0.1310\\
0.35 & 0.9973 & 0.9901  & 0.9 & 0.0857 & 0.0798\\
0.4  & 0.8263 & 0.8108  & 1.0 & 0.0542 & 0.0482\\
0.5  & 0.5460 & 0.5330  & 1.5 & 0.0067 & 0.0059\\
0.6  & 0.3488 & 0.3405  & 2.0 & 0.0012 & 0.0010\\
0.7  & 0.2190 & 0.2088  & --  &  --    & --    \\
\hline
\end{tabular}
\end{center}
\end{table}

\begin{table}[ht]
\begin{center}
\caption{Form Factor Values of $B$ meson in finite and infinite mass limit.}\label{cross}
\begin{tabular}{|c|c|c||c|c|c|}
  \hline
 $Q^2$ & $F(Q^2)$ & $F(Q^2)|_{\infty}$ &  $Q^2$ & $F(Q^2)$& $F(Q^2)|_{\infty}$  \\
\hline \hline
0    &  1     & 1       & 0.7 & 0.7624 & 0.7584\\
0.1  & 0.9957 & 0.9899  & 0.8 & 0.7041 & 0.7005\\
0.2  & 0.9786 & 0.9730  & 0.9 & 0.6448 & 0.6402\\
0.3  & 0.9510 & 0.9456  & 1.0 & 0.5859 & 0.5731\\
0.4  & 0.9140 & 0.9089  & 1.5 & 0.3301 & 0.3226\\
0.5  & 0.8691 & 0.8640  & 2.0 & 0.1681 & 0.1611\\
0.6  & 0.8179 & 0.8136  & 2.5 & 0.0822 & 0.0784\\
\hline
\end{tabular}
\end{center}
\end{table}

\begin{table}[ht]
\begin{center}
\caption{Charge radii of mesons in finite and infinite mass limit.}\label{cross}
\begin{tabular}{|c|c|c|}
  \hline
 $Meson$ & $<r^2>$ & $R=\frac{<r^2>}{<r^2>_\infty}$  \\
\hline \hline
$D_0$    & -0.52 & 0.9824  \\
$D^+$    & 0.27 & 0.9660  \\
$D_s^+$  & 0.21 & 0.6758  \\
$B^+$    & 0.54 & 1.9977  \\
$B_s^0$  & -0.32 & 0.9517  \\
$B_c^+$  & -0.24 & 1.7693  \\
\hline
\end{tabular}
\end{center}
\end{table}

\begin{table}[ht]
\begin{center}
\caption{Decay constants of $D$ and $B$ mesons(in MeV).}\label{cross}
\begin{tabular}{|c|c|c||c|c|c|}
  \hline
 $D Meson$ & Mass & Decay Constant&  $B Meson$ & Mass & Decay Constant\\
\hline \hline
$D[D^+,D^-]$                 & 1869.6 & 350.66  & $B[B^+,B^-]$                  & 5279.1 & 215.20  \\
$D^0[D^0,\overline{D^0}]$    & 1864.8 & 351.11  & $B^0[B^0,\overline{B^0}]$     & 5279.5 & 215.17 \\
$D_s[D_s^+,D_s^-]$           & 1968.5 &  341.74 & $B_s[B_s^0,\overline{B_s^0}]$ & 5366.3 & 213.45 \\
--                           &  --    & --      & $B_c[B_c^+,B_c^-]$            & 6276.0 & 197.37 \\
\hline
\end{tabular}
\end{center}
\end{table}

\begin{table}[ht]
\begin{center}
\caption{Some theoretical estimates of weak decay constant $f_{D_s}$ (in MeV).}\label{cross}
\begin{tabular}{|r|r|r|r|}
  \hline
 Potential Model & Bag Model & Sum rules &  Lattice\\
\hline \hline
260 [44]               & 166 [51] & 232 [52]       &  $215\pm17$ [56] \\
149  [45]              &          & $276\pm13$ [53] &  $157\pm11$ [57] \\
210   [46]             &          & $218\pm20$ [54] &  $234\pm72$ [58] \\
380-590 [47]           &          & $200\pm15$ [55] &  280    [59]    \\
356[48]                &          &                 &  $209\pm18 $  [60]\\
199 [49]               &          &                 &             \\
$290\pm20$ [50]        &          &                 &             \\
\hline
\end{tabular}
\end{center}
\end{table}

\begin{table}[ht]
\begin{center}
\caption{Some theoretical estimates of weak decay constant $f_{D}$ (in MeV).}\label{cross}
\begin{tabular}{|r|r|r|}
  \hline
 Potential Model & Sum rules &  Lattice\\
\hline \hline
112-141 [61]               & 290 [51]           & $198\pm17$ [60]\\
200  [44]                  & $170\pm30$  [63]   &  \\
139   [45]                 &                  &  \\
112-137 [46]               &                  &    \\
360-580[47]                &                  &    \\
281 [48]                   &                  &   \\
150 [62]                   &                  &   \\
\hline
\end{tabular}
\end{center}
\end{table}

\begin{table}[ht]
\begin{center}
\caption{Some theoretical estimates of weak decay constant $f_{B}$ (in MeV).}\label{cross}
\begin{tabular}{|r|r|r|r|}
  \hline
 Potential Model  & Sum rules & Factorisation & Lattice\\
\hline \hline
120 [44]               & 290 [52]              &  $150\pm50$ [67] & $366\pm22\pm55$ [60] \\
93  [45]               & $190\pm50$ [64]         &                 & $205\pm40$ [68]       \\
$75-114$   [46]          & $200\pm35$ [65]         &                 &  $310\pm25\pm50$ [69] \\
260-300 [47]           & $170\pm20$ [66]         &                 &  $233\pm42$ [69]    \\
229[48]                &  140 [53]               &                 &                     \\
$<100$ [62]            &                       &                 &                     \\
\hline
\end{tabular}
\end{center}
\end{table}

\begin{figure}
    \centering
    {        \includegraphics[width=3.0in]{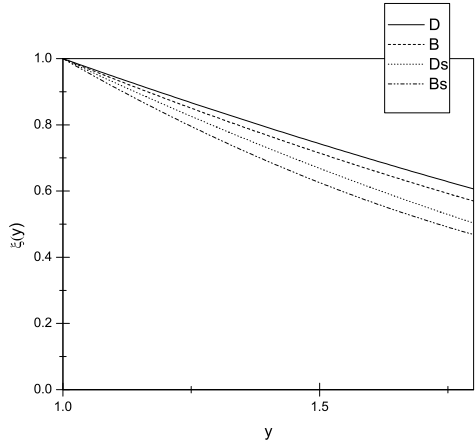}
        \label{fig}
    }
\caption{Variation of IWF with y}
\label{fig:sample_subfigures}
\end{figure}

\begin{figure}
    \centering
    \subfigure[ $F(Q^2)$ vs $Q^2$ for D meson]
    {
        \includegraphics[width=3.0in]{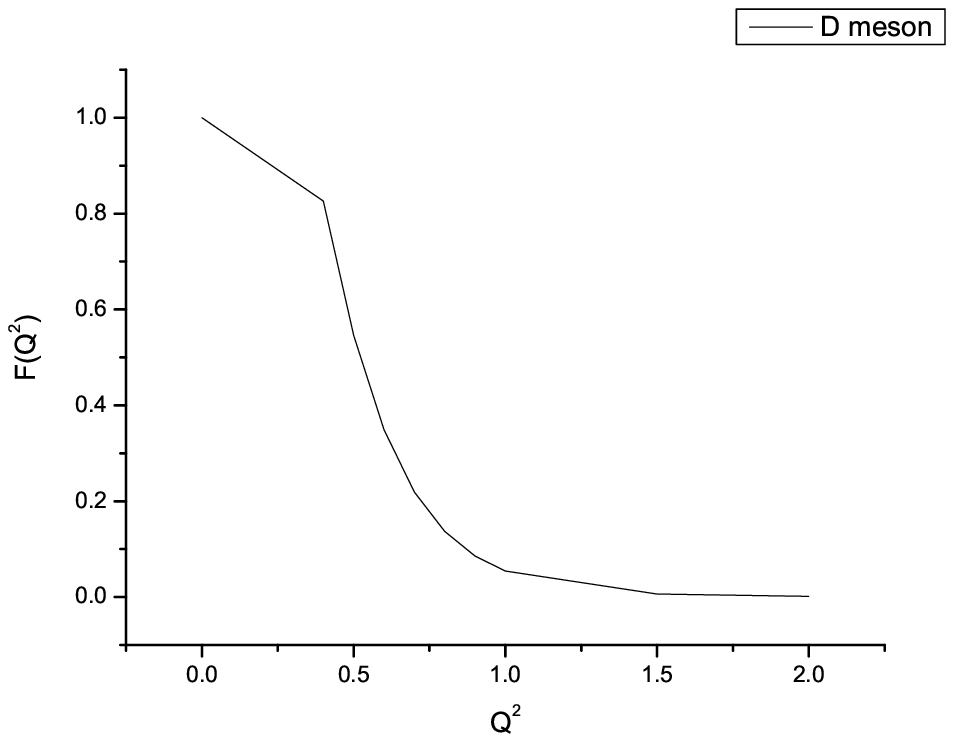}
        \label{fig:first_sub}
    }
        \subfigure[ $F(Q^2)$ vs $Q^2$ for B meson]
    {
        \includegraphics[width=3.0in]{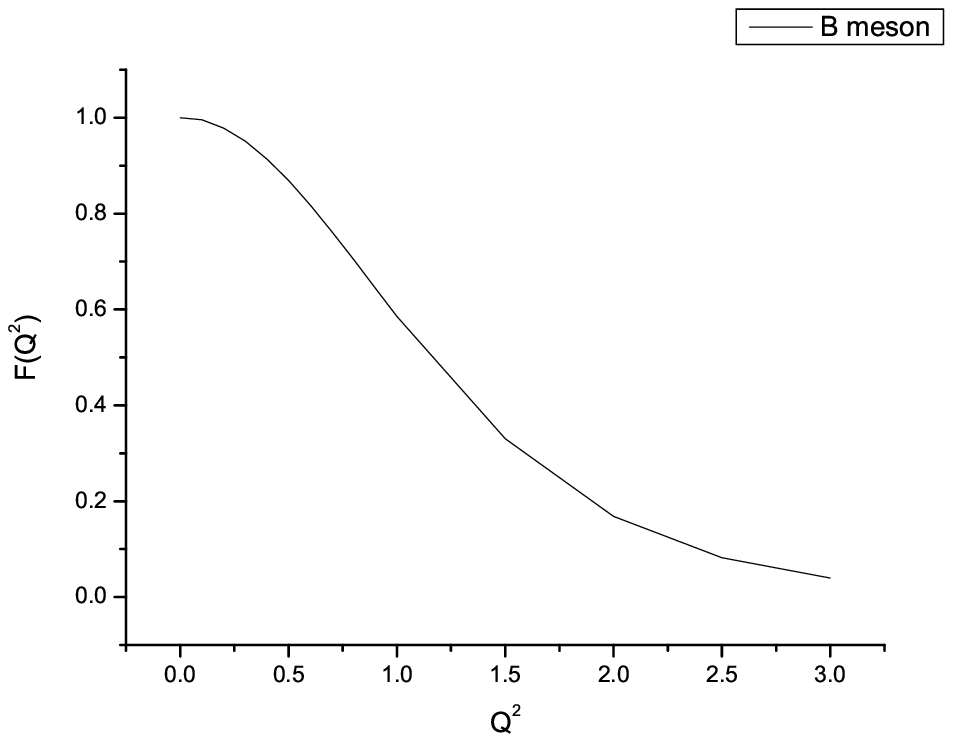}
        \label{fig:second_sub}
    }
        \caption{Variation of form factor with $Q^2$ for $D$ and $B$ mesons}
    \label{fig:sample_subfigures}
\end{figure}

\begin{figure}
    \centering
    {        \includegraphics[width=3.0in]{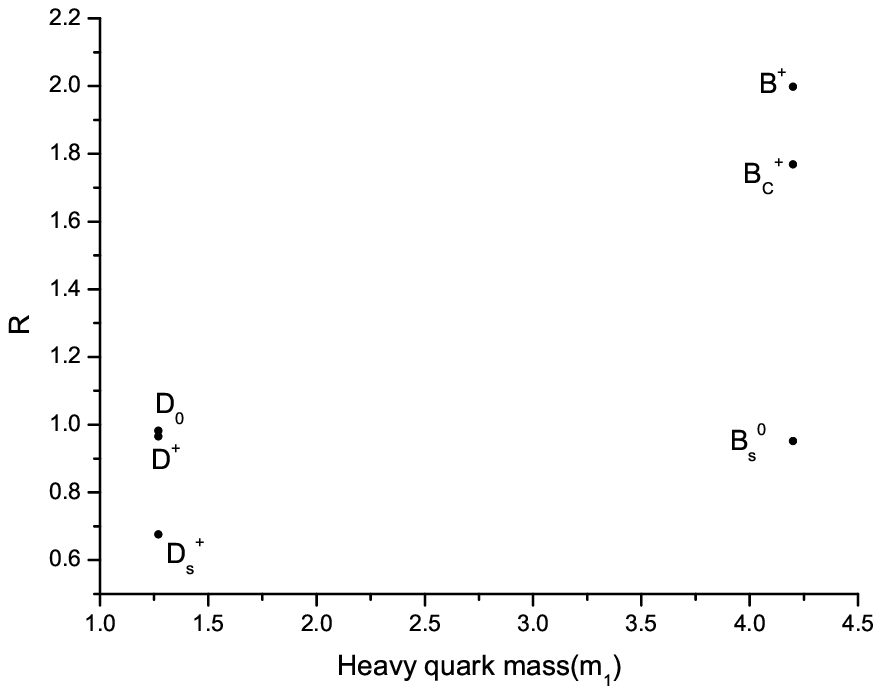}
        \label{fig}
    }
\caption{Ratio $R=\frac{<r^2>}{<r^2>_\infty}$ vs heavy quark mass $(m_1)$}
\label{fig:sample_subfigures}
\end{figure}

\end{document}